\documentclass[reprint,superscriptaddress,showkeys,floats,floatfix]{revtex4-1}
\usepackage{graphicx}
\usepackage{color}
\usepackage{amsmath}

\begin{document}

\title{Pauli Spin Blockade in a Highly Tunable Silicon Double Quantum Dot}

\author{N. S. Lai}
\email[Electronic mail: ]{ns.lai@student.unsw.edu.au}
\affiliation{ARC Centre of Excellence for Quantum Computation and Communication Technology, School of Electrical Engineering \& Telecommunications, The University of New South Wales, Sydney 2052, Australia}
\author{W. H. Lim}
\affiliation{ARC Centre of Excellence for Quantum Computation and Communication Technology, School of Electrical Engineering \& Telecommunications, The University of New South Wales, Sydney 2052, Australia}
\author{C. H. Yang}
\affiliation{ARC Centre of Excellence for Quantum Computation and Communication Technology, School of Electrical Engineering \& Telecommunications, The University of New South Wales, Sydney 2052, Australia}
\author{F. A. Zwanenburg}
\affiliation{ARC Centre of Excellence for Quantum Computation and Communication Technology, School of Electrical Engineering \& Telecommunications, The University of New South Wales, Sydney 2052, Australia}
\author{W. A. Coish}
\affiliation{Department of Physics, McGill University, Montreal, Quebec H3A 2T8, Canada}
\author{F. Qassemi}
\affiliation{Institute for Quantum Computing and Department of Physics and Astronomy, University of Waterloo, Waterloo, Ontario N2L 3G1, Canada}
\author{A. Morello}
\affiliation{ARC Centre of Excellence for Quantum Computation and Communication Technology, School of Electrical Engineering \& Telecommunications, The University of New South Wales, Sydney 2052, Australia}
\author{A. S. Dzurak}
\affiliation{ARC Centre of Excellence for Quantum Computation and Communication Technology, School of Electrical Engineering \& Telecommunications, The University of New South Wales, Sydney 2052, Australia}


\date{\today}

\begin{abstract}

Double quantum dots are convenient solid-state platforms to encode quantum information. Two-electron spin states can be conveniently detected and manipulated using strong quantum selection rules based on the Pauli exclusion principle, leading to the well-know Pauli spin blockade of electron transport for triplet states. Coherent spin states would be optimally preserved in an environment free of nuclear spins, which is achievable in silicon by isotopic purification. Here we report on a deliberately engineered, gate-defined silicon metal-oxide-semiconductor double quantum dot system. The electron occupancy of each dot and the inter-dot tunnel coupling are independently tunable by electrostatic gates. At weak inter-dot coupling we clearly observe Pauli spin blockade and measure a large intra-dot singlet-triplet splitting $>$ 1 meV. The leakage current in spin blockade has a peculiar magnetic field dependence, unrelated to electron-nuclear effects and consistent with the effect of spin-flip cotunneling processes. The results obtained here provide excellent prospects for realizing singlet-triplet qubits in silicon.

\end{abstract}



\maketitle

Gate-defined semiconductor quantum dots enable the confinement and manipulation of individual electrons and their spin~\cite{Hanson2007}. Most of the relevant parameters -- electron filling, energy splittings, spin states, exchange interaction -- can be tuned \emph{in situ} by electric and magnetic fields. Because of this exquisite level of control, quantum dots are being investigated as candidate systems for spin-based quantum information processing~\cite{Loss1998}. In group III-V semiconductors such as GaAs, the development of highly tunable double quantum dots has allowed the study of both single-electron and two-electron spin dynamics~\cite{Petta2005,Koppens2006,Ono2002,Koppens2005,Johnson2005}. However, the nuclear spins always present in these materials produce strong decoherence of the electron spin degree of freedom and result in phase coherence times $T_2$ of below 1 ms~\cite{Barthel2010,Bluhm2010}. Conversely, group-IV semiconductors such as silicon, silicon-germanium and carbon can be isotopically purified, leaving only spinless isotopes. The weak spin-orbit coupling~\cite{Tahan2002} and the absence of piezoelectric electron-phonon coupling~\cite{Fedichkin2004} allow for extremely long spin relaxation times $T_1$ of order seconds, as already demonstrated in several experiments~\cite{Xiao2010,Morello2010,Simmons2011}. The phase coherence times have not been measured yet, but they are expected to reach $\sim 1$~s as well, in highly purified $^{28}$Si substrates with low background doping concentration~\cite{Witzel2010}.

\begin{figure}
\includegraphics[width=8cm]{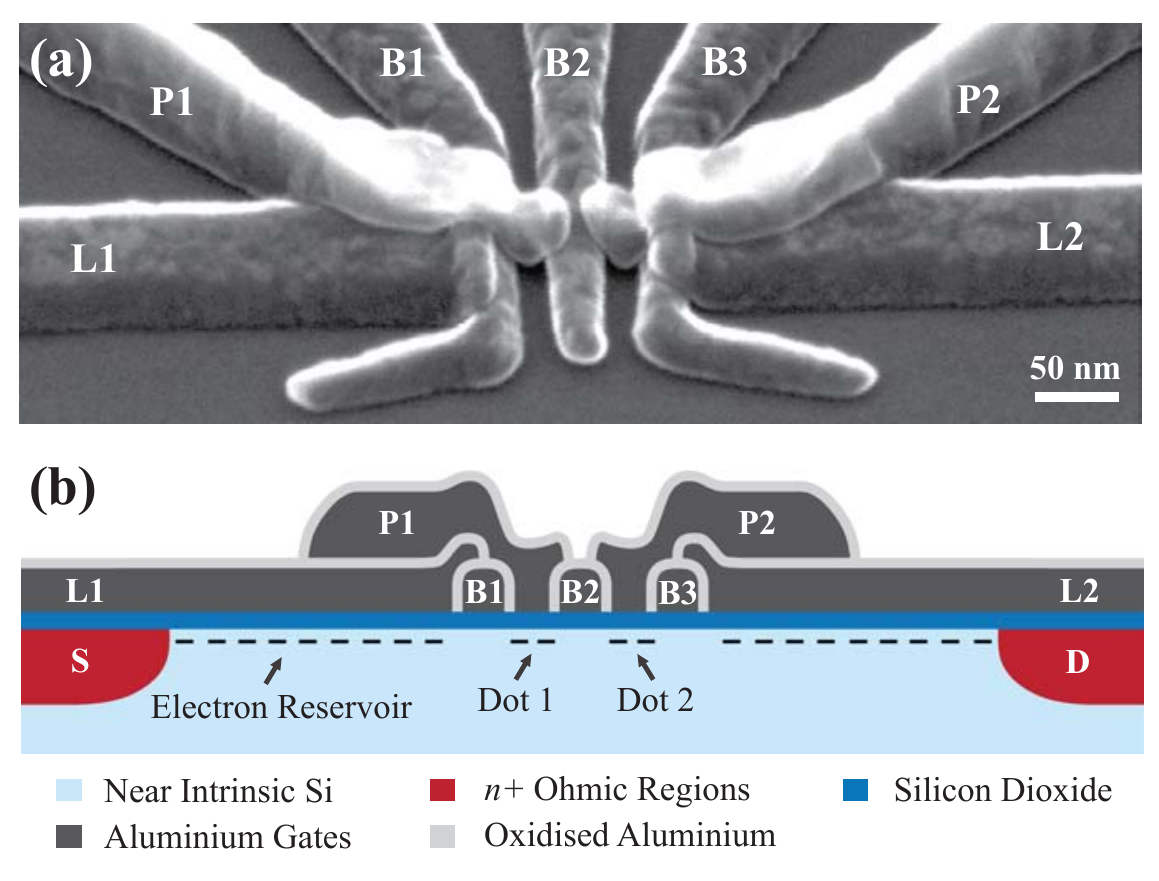}
\caption{{\bf SEM and schematic view of the device}. (a) Scanning electron micrograph of a device identical to that measured. (b) (Not to scale) Schematic cross-section view of the Si MOS double quantum dot. The architecture is defined by B1, B2 and B3 (barrier gates), L1 and L2 (lead gates), and P1 and P2 (plunger gates). The gates are separated by an Al$_2$O$_3$ layer (light gray). Positive voltages applied to the lead and plunger gates induce an electron layer (black dashes) underneath the SiO$_2$. By tuning the barrier gates, Dot 1 and Dot 2 are formed. The coupling of the dots is adjusted using the middle barrier (B2). The regions coloured with red are the $n^+$ source (S) and drain (D) contacts formed via diffused phosphorus.} \label{fig1}
\end{figure}

A widely successful method to observe and control spin phenomena in quantum dots~\cite{Hanson2007} consists of defining a double quantum dot in a series configuration and tuning the potentials such that sequential electron transport requires a stage where two electrons must occupy the same dot. The eigenstates of a two-electron system are singlet and triplet spin states, separated by an energy splitting $\Delta_{\rm ST}$ which can be large in tightly confined dots. The electron transport then becomes spin-dependent and can be blocked altogether when the two-electron system forms a triplet state~\cite{Ono2002, Fransson2006}. This phenomenon, known as Pauli spin blockade, has been extensively exploited to investigate the coherence of single-spin~\cite{Koppens2006} and two-spin states~\cite{Petta2005} in GaAs and InAs~\cite{Nadj-perge2010B} quantum dots. Therefore, observing and controlling spin blockade in silicon is a key milestone to unravel the full potential of highly coherent spin qubits. Preliminary success has been obtained in Si~\cite{Liu2008} and SiGe~\cite{Shaji2008} devices, but in each case the double dot system under study resulted from local variations in the potential of a lithographically-defined single dot, making it difficult to control individual dot occupancies or inter-dot coupling. Spin-based quantum dot qubits require exquisite control of these parameters, so a highly tunable double-dot system in silicon is essential. For singlet-triplet qubits in multivalley semiconductors it is also crucial to ensure that a large valley-orbit splitting is present, to avoid the lifting of Pauli blockade due to valley degeneracy \cite{Palyi2009,Culcer2010}.

Here we present an engineered silicon double quantum dot which shows excellent tunability and robust charge stability over a wide range of electron occupancy (\emph{m, n}). The silicon metal-oxide-semiconductor (MOS) structure utilizes an Al-Al$_2$O$_3$-Al multi-gate stack that enables very small dots to be defined, each with independent gate control, together with gate-tunable inter-dot coupling. Such multi-gate stacks have previously been used to construct single Si quantum dots with the ability to achieve single electron occupancy~\cite{Lim2009}. The double dot presented here exhibits spin blockade in the few-electron regime, from which we are able to extract a large singlet$-$triplet energy splitting and also investigate a new mechanism of singlet$-$triplet mixing in the weak-coupling regime.

\section{Results}

{\bf Device architecture}. Figure~\ref{fig1} shows a scanning electron micrograph (SEM) and cross-sectional schematic of the device, which incorporates 7 independently controlled aluminium gates. When a positive bias is applied to the lead gates (L1 and L2) an accumulation layer of electrons is induced under the thin SiO$_2$, to form the source and drain reservoirs for the double dot system. A positive voltage on the plunger gate P1 (P2) causes electrons to accumulate in Dot 1 (Dot 2). Independent biasing of P1 and P2 provides direct control of the double-dot electron occupancy (\emph{m, n}). The tunnel barriers between the two dots and the reservoirs are controlled using the barrier gates: B1, B2 and B3. The middle barrier gate B2 determines the inter-dot tunnel coupling. The electrochemical potentials of the coupled dots can also be easily tuned to be in resonance with those of the source and drain reservoirs. As shown in Fig.~\ref{fig1}(b), gates L1 and L2 extend over the source and drain $n^+$ contacts, and also overlap gates B1 and B3. The upper-layer gates (P1 and P2) are patterned on top of the lead and barrier gates. The lithographic size of the dots is defined by the distance between adjacent barrier gates ($\sim$30~nm) and the width of the plunger gates ($\sim$50~nm), as shown in Fig.~\ref{fig1}(a).

\begin{figure}[t]
\includegraphics[width=8.5cm]{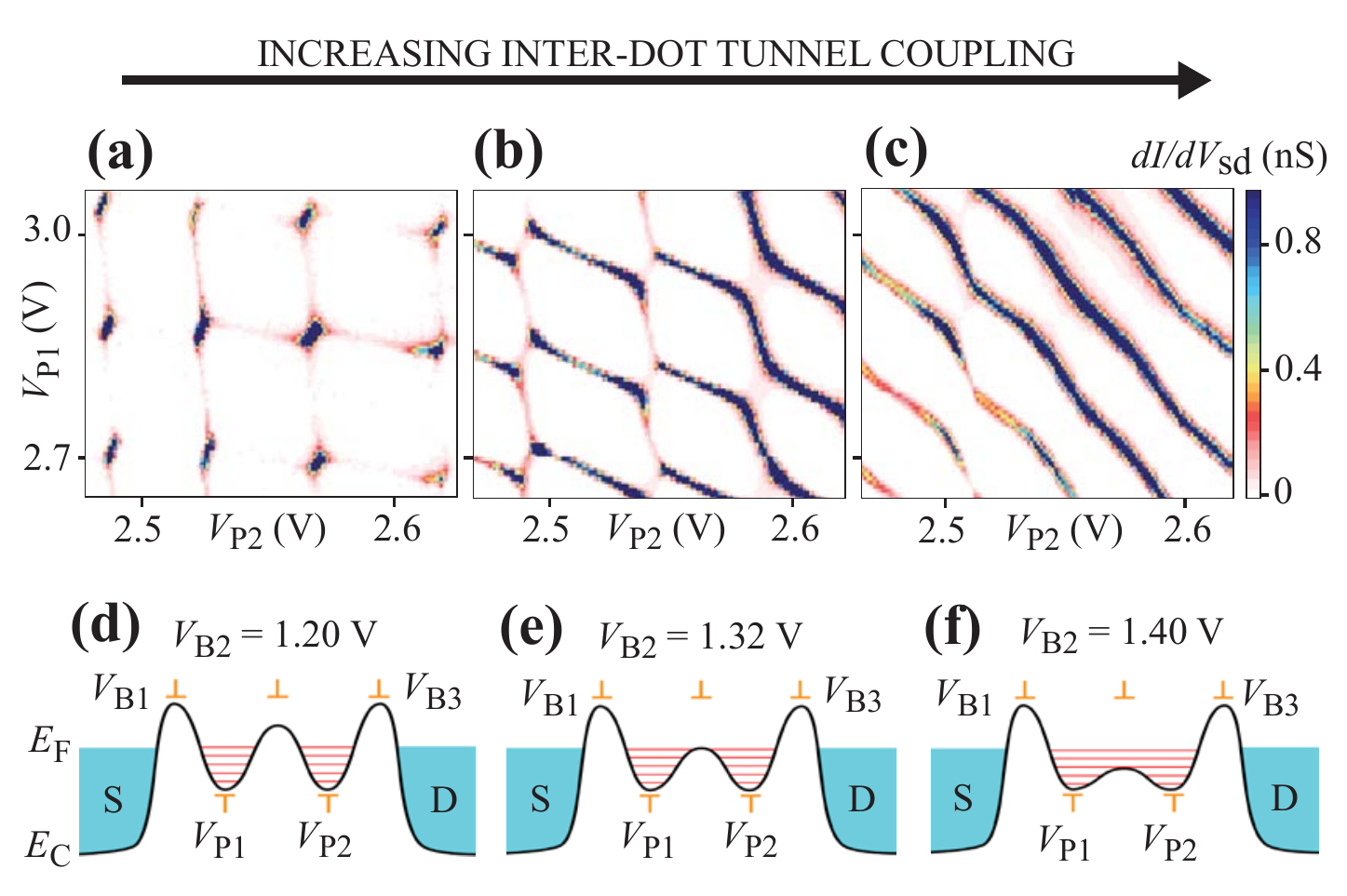}
\caption{{\bf Characteristics at different inter-dot tunnel coupling}. Measured stability diagrams and energy landscape of the double dot system ranging from weak to strong inter-dot tunnel coupling (a)$-$(c) and (d)$-$(f) respectively, for $V_{\rm L1}$ = $V_{\rm L2}$ = 3.0~V, $V_{\rm B1}$ = 0.76~V, $V_{\rm B3}$ = 1.0~V and $V_{\rm SD}$ = 0. From lower to higher $V_{\rm B2}$, the tunnel barrier height decreases resulting in stronger inter-dot tunnel coupling. (a) A checker box pattern, (b) honeycomb pattern and (c) diagonal parallel lines indicate that the two dots merge into a single dot as the coupling is increased~\cite{Wiel2002}.}
\label{fig2}
\end{figure}

{\bf Inter-dot tunnel coupling tunability}. Figure~\ref{fig2} shows the measured differential conductance of the device as a function of the plunger gate voltages, $V_{\rm P1}$ and $V_{\rm P2}$, with all other gate voltages held constant, together with sketches of the energy landscape of the double dot. The charge-stability maps moving from Fig.~\ref{fig2}(a) to~\ref{fig2}(c) clearly show the effects of an increasing inter-dot coupling as the middle barrier-gate voltage $V_{\rm B2}$ is increased, lowering the tunnel barrier between the dots. Fig.~\ref{fig2}(b) shows the characteristic honeycomb-shaped stability map representing intermediate inter-dot coupling~\cite{Wiel2002}, obtained at $V_{\rm B2}$ = 1.32 V. At lower middle barrier-gate voltage, $V_{\rm B2}$ = 1.20 V, we observe a checker-box shaped map [Fig.~\ref{fig2}(a)], since the middle barrier is opaque enough to almost completely decouple the two dots. In contrast, the stability map in Fig.~\ref{fig2}(c) shows the formation of diagonal parallel lines at $V_{\rm B2}$ = 1.40 V. Here the two dots effectively merge into a single dot due to the lowering of the middle barrier [Fig.~\ref{fig2}(f)]. The transport measurements shown here do not allow a precise determination of the electron occupancy (\emph{m, n}) in the dots, since it is possible that electrons remain in the dots even when $I_{\rm SD}$ is immeasurably small. For the regime plotted in Fig.~\ref{fig2} there were at least 10 electrons in each dot, based on our measurement of Coulomb peaks as we further depleted the system. An absolute measurement of dot occupancy would require integration of a charge sensor into the system~\cite{Johnson2005}. These results nevertheless demonstrate that the multi-gated structure provides excellent tunability of coupling while maintaining charge stability over a wide range of electron occupancy.

\begin{figure}[t]
\includegraphics[width=8.5cm]{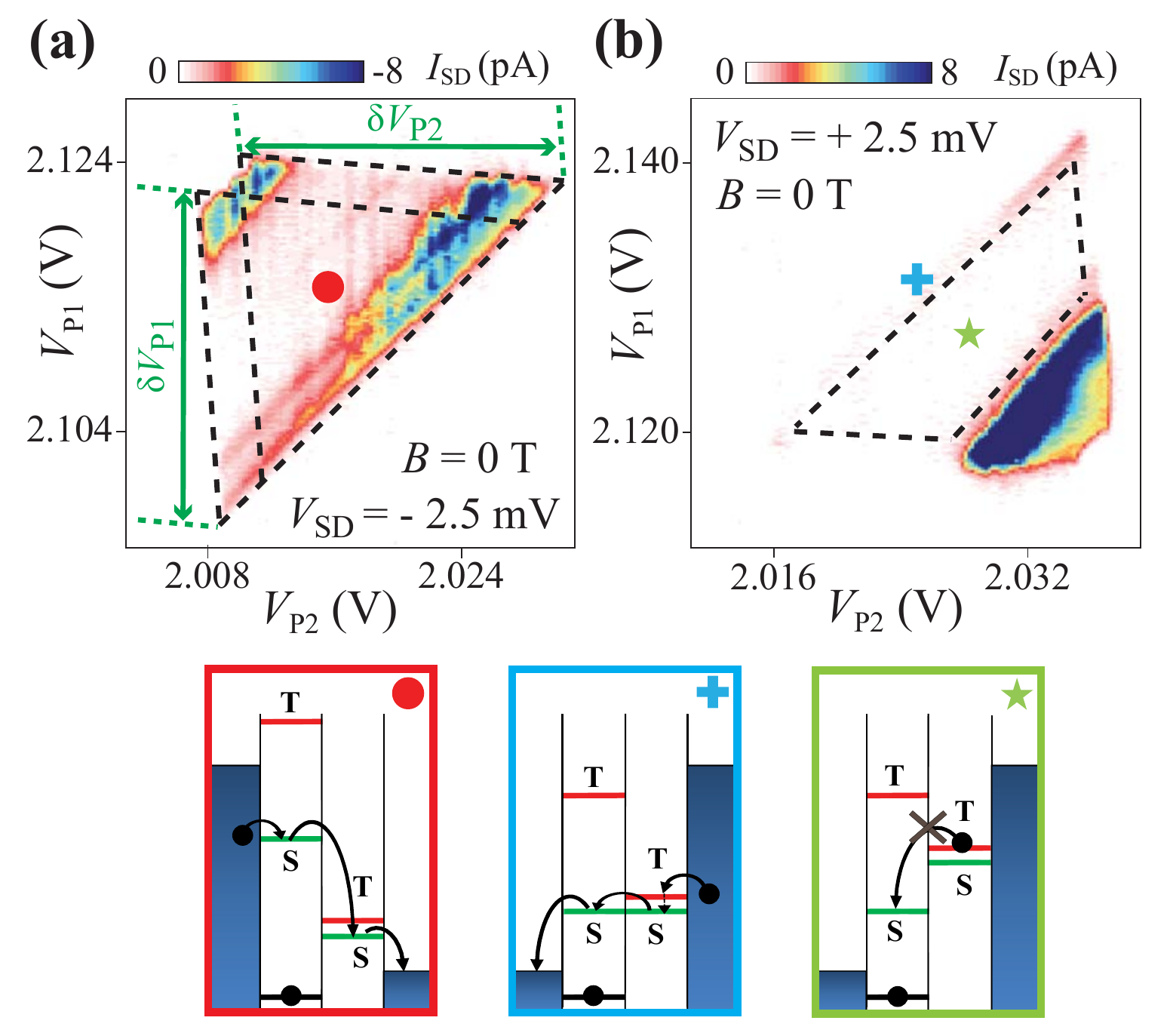}
\caption{{\bf Pauli spin blockade at weakly coupled regime}. Current $I_{\textrm{SD}}$ as a function of $V_{\rm P1}$ and $V_{\rm P2}$ for $B$ = 0 T. The lead and barrier gate voltages were fixed at $V_{\rm L1}$ = $V_{\rm L2}$ = 3.2~V, $V_{\rm B1}$ = 0.656~V, $V_{\rm B2}$ = 1.176~V and $V_{\rm B3}$ = 0.940~V throughout the experiment. (a) For $V_{\rm SD}$ = $-$2.5 mV, the ground state and excited states of a full bias triangle are shown. The current flows freely at the S(0,2)$-$S(1,1) transition as illustrated in the box marked by red dot. (b) The same configuration at $V_{\rm SD}$ = +2.5 mV, the current between the singlet and triplet states is fully suppressed by spin blockade (green star box) except on the bottom (blue cross box) of the bias triangle. The blue cross box shows how a leakage current arises the Pauli spin blockade region.}
\label{fig3}
\end{figure}

{\bf Capacitances and charging energies}. Application of a DC source-drain bias $V_{\rm SD}$ causes the triple-points in the weakly-coupled regime [Fig.~\ref{fig2}(a)] to extend to form triangular shaped conducting regions [Fig.~\ref{fig3}(a)] from which the energy scales of the double dot system can be determined~\cite{Wiel2002}. From a triangle pair, we extract the conversion factors between the gate voltages and energy to be $\alpha_1$ = $eV_{\rm SD}$/$\delta V_{\rm P1} = 0.089e$ and $\alpha_2 = eV_{\rm SD} / \delta V_{\rm P2} = 0.132e$, where $\delta V_{\rm P1}$ and $\delta V_{\rm P2}$ are the lengths of the triangle edges, as shown in Fig.~\ref{fig3}(a). The total capacitances of Dot-1 and Dot-2 can then be calculated~\cite{Wiel2002}, giving $C_1$ = 16.3 aF and $C_2$ = 14.5 aF. The charging energies of the two dots are then $E_{\rm C,1} = e^2 / C_1$ = 9.8 meV and $E_{\rm C,2} = e^2/C_2$ = 11 meV, indicating that the left dot is slightly larger than the right dot.

{\bf Pauli spin blockade}. Figure~\ref{fig3} shows the current $I_{\rm SD}$ through the double dot as a function of the two plunger gate voltages when measured with both positive [Fig.~\ref{fig3}(a)] and negative [Fig.~\ref{fig3}(b)] source-drain biases. Here we observe a suppression of current at one bias polarity, the characteristic signature of Pauli spin blockade~\cite{Koppens2005,Johnson2005}. At $V_{\rm SD}$ = $-$2.5 mV we observe a pair of overlapping full bias triangles, as shown in Fig.~\ref{fig3}(a). Resonant transport through the ground state and the excited states in the double dot occurs when the states within the dots are exactly aligned, leading to peaks in the current which appear as straight lines parallel to the triangle base in Fig.~\ref{fig3}(a). The non-resonant background current level at the centre of the triangle is attributed to inelastic tunneling. The non-zero current throughout the triangular region indicates that electrons from the reservoir can tunnel freely from the S(0,2) singlet state to the S(1,1) singlet state, as depicted in the cartoon (red box in Fig.~\ref{fig3}). Note that here we define (\emph{m, n}) as the \emph{effective} electron occupancy~\cite{Liu2008}, while the \emph{true} electron occupancy is ($m$+$m_0$, $n$+$n_0$). The Pauli blockade expected for two-electron singlet and triplet states occurs when the total electron spin of each dot is zero in the ($m_0$, $n_0$) state.

At the complementary positive bias of $V_{\rm SD}$ = +2.5 mV we observe strong current suppression in the region bounded by the dashed lines in Fig.~\ref{fig3}(b). The suppression arises because the transition from T(1,1) to S(0,2) is forbidden by spin conservation during electron tunneling. Once the T(1,1) triplet state is occupied, further current flow is blocked until the electron spin on the right dot reverses its orientation via a relaxation process (green star box in Fig.~\ref{fig3})~\cite{Koppens2005,Johnson2005}.

\begin{figure}
\includegraphics[width=8.5cm]{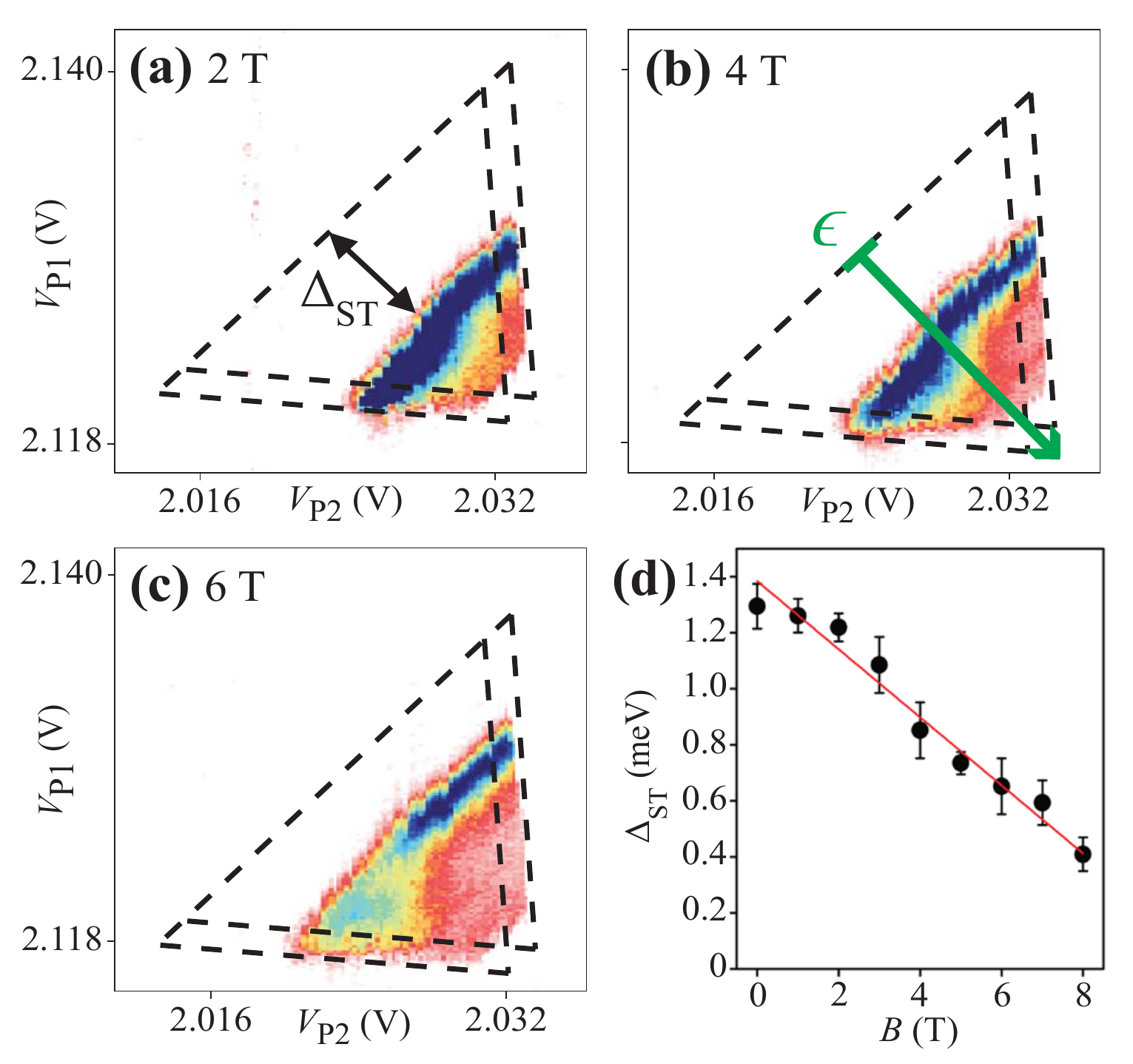}
\caption{{\bf Singlet-triplet splitting}. (a)$-$(c) DC measurements of the triangle pair analysed in Fig.~\ref{fig3}, at $V_{\rm SD}$ = +2.5~mV, for different in-plane magnetic fields, $B$ (scale bar same as Fig.~\ref{fig3}(b)). The singlet$-$triplet splitting, $\Delta_{\rm ST}$, is defined by the triplet and singlet state of (0,2) as depicted in (a). As the magnetic field increases, $\Delta_{\rm ST}$ decreases along the detuning axis of the triangle [labeled $\epsilon$ in (b)]. (d) The energy spacing $\Delta_{\rm ST}$ as a function of in-plane magnetic field $B$. $\Delta_{\rm ST}$ decreases at a rate $\sim$0.12 meV/T and is expected to approach zero at 11.3 T. From the linear fit (red line) through $\Delta_{\rm ST}$, the $g$-factor is 2.1 $\pm$ 0.2.}
\label{fig4}
\end{figure}

{\bf Singlet-triplet splitting}. In a magnetic field $B$ there are four accessible spin states: the singlet S; and three triplets T$_-$, T$_0$ and T$_+$, corresponding to $S_{\rm Z}$ = $-$1, 0, +1. The singlet$-$triplet splitting $\Delta_{\rm ST}$ is the energy difference between the blockaded ground state S(0,2) and the excited state T$_-$(0,2)~\cite{Johnson2005,Liu2008}. Here we study $\Delta_{\rm ST}$ as a function of $B$, applied parallel to the substrate, by measuring spin blockade at a positive bias. Figures~\ref{fig4}(a$-$c) show the bias triangles in the spin blockade regime at increasing magnetic fields $B$ = 2, 4 and 6 T, with the splitting $\Delta_{\rm ST}$ marked in Fig.~\ref{fig4}(a). The measured splitting $\Delta_{\rm ST}$ decreases linearly with increasing $B$ [Fig.~\ref{fig4}(d)], as expected, since the triplet states split linearly by the Zeeman energy, $E_{\rm Z}$ = $\pm S_{\rm Z}|g|\mu_BB$, where $\mu_B$ is the Bohr magneton and $S_{\rm Z}$ is $-$1, 0, +1. A linear fit through $\Delta_{\rm ST}(B)$ yields a Land\'{e} $g$-factor of 2.1 $\pm$ 0.2, consistent with electrons in silicon.

We observe an exceptionally large value of the (0,2) singlet-triplet splitting at $B=0$, $\Delta_{\rm ST} \approx 1.4$~meV. This result is striking because it implies that the nearest valley-orbit state must be at least 1.4 meV above the ground state. The first excited valley-orbit state should be a combination of the $\pm z$ valleys. It would lift the spin blockade \cite{Palyi2009,Culcer2010}, and show no remarkable energy shift in a magnetic field, in contrast with our observations. Therefore, such a state must lie above the triplet state we observe in Fig.~\ref{fig4}. The ability of our structures to generate such a large valley-orbit splitting removes a major concern on the realizability of singlet-triplet qubits in a multivalley material such as silicon.

\begin{figure}[t]
\includegraphics[width=8.5cm]{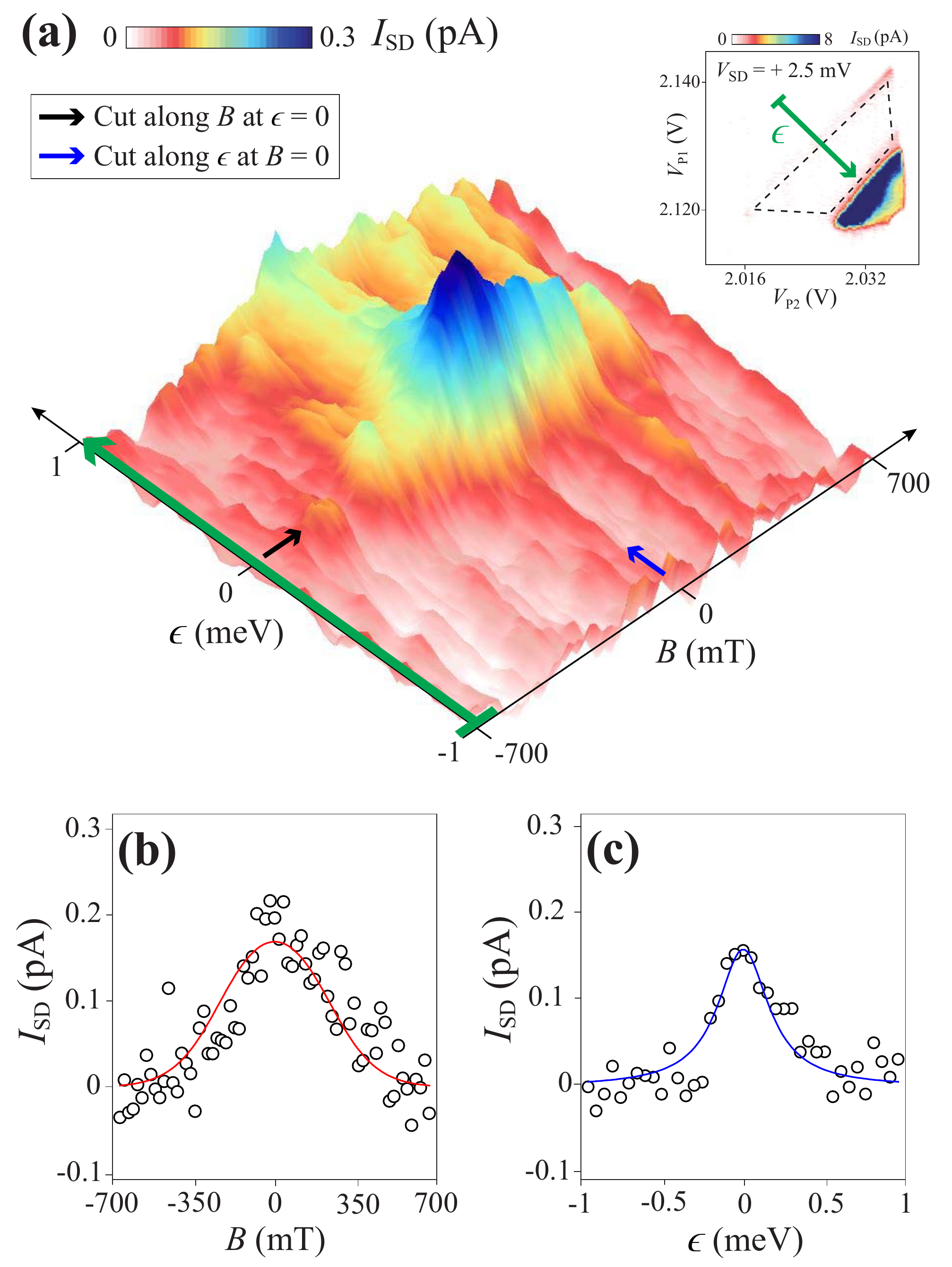}
\caption{{\bf Spin-flip cotunneling in Pauli spin blockade regime}. (a) A surface plot of leakage current through spin blockade as a function of energy detuning $\epsilon$ and magnetic field $B$, with gates settings as in Fig~\ref{fig3}(b). (b) Cut along $B$ at $\epsilon = 0$ energy detuning axis (black arrow) while (c) Cut along $\epsilon$ at $B = 0$ field (blue arrow). Fits of experimental data with spin-flip cotunneling relation give $\Gamma_{\mathrm{D}}$ = 34 $\mu$eV, $t$ = 0.5 $\mu$eV and $T$ = 115 mK.} \label{fig5}
\end{figure}

{\bf Leakage current in blockade regime}. If some mechanism exists to mix the singlet and triplet states or to induce transitions between them, then the spin blockade can be lifted, leading to a measurable leakage current~\cite{Koppens2005}. As shown in the blue cross box in Figure~\ref{fig3}, transitions from T(1,1) to S(1,1) can lift the blockade, allowing electrons to transit the double dot until the next triplet is loaded, resulting in a non-zero time-averaged leakage current. Fig.~\ref{fig5}(a) shows the surface plot of the leakage current $I_{\rm SD}$ as a function of both detuning $\epsilon$ and magnetic field $B$, while Figs.~\ref{fig5}(b) and~\ref{fig5}(c) show line traces of $I_{\rm SD}$ as a function of $B$ at zero detuning and $I_{\rm SD}$ as a function of $\epsilon$ at zero magnetic field, respectively. We find that the leakage current has a maximum at $B \approx 0$ and falls to zero at $|B| \sim 700$~mT.

\section{Discussion}

The suppression of leakage current by an applied magnetic field has been observed in GaAs double quantum dots (DQDs)~\cite{Koppens2005} and attributed to the effect of hyperfine coupling between the electron spins and the surrounding bath of nuclear spins. In that case the width $\delta B$ of $I_{\rm SD}(B)$ yields the average strength of the hyperfine field. For an unpolarized nuclear spin bath $\delta B \approx \delta B_{\rm max} / \sqrt{N}$, where $\delta B_{\rm max}$ is the hyperfine field assuming fully polarized nuclei and $N$ is the number of nuclei overlapping with the electron wave function. For a typical GaAs dot overlapping with $\sim 10^6 - 10^7$ nuclei, $\delta B_{\rm max} \sim 6$~T $\Rightarrow \delta B \sim 2 - 6$~mT~\cite{Koppens2005,Johnson2005,Merkulov2002}. In natural silicon, however, the hyperfine interaction is much smaller than in GaAs, with $\delta B_{\rm max} \approx 1.9$~mT~\cite{Assali2010}. Therefore, hyperfine coupling can be ruled out as a mechanism for the leakage current here.

An alternative mechanism for a transition from triplet to singlet has been recently proposed, where the spin flip is caused by inelastic cotunneling~\cite{Qassemi2009}. The spin-flip rates due to cotunneling from the spin-polarized triplet states, $T_\pm (1,1)$, are exponentially suppressed when the Zeeman energy is large compared to the thermal broadening of the electron states in the leads (i.e., for $g\mu_\mathrm{B}B>k_\mathrm{B} T$, where $T$ is the electron temperature and $B$ is the applied magnetic field). A rate-equation analysis accounting for the energy dependence of the spin-flip cotunneling rates \cite{Qassemi2009} then gives a simple form in the limit of weak inter-dot tunneling $t$ and weak cotunneling $W_\mathrm{cot}^0$ compared to the tunnel rates $\Gamma_{\mathrm{S},\mathrm{D}}$ between a dot and its nearby source or drain lead ($\sqrt{2}t < k_\mathrm{B} T$, $W_\mathrm{cot}^0\ll \Gamma_{\mathrm{S},\mathrm{D}}$) \cite{Coish2011}:
\begin{equation}\label{eq:IvsB}
 I_\mathrm{SD}(B) = e\frac{4}{3}\frac{3 W_\mathrm{cot}^0/2}{1+\frac{k_\mathrm{B}T}{2 g\mu_\mathrm{B} B}\sinh\left(\frac{g\mu_\mathrm{B} B}{k_\mathrm{B} T}\right)},\quad \epsilon=0.
\end{equation}
Here, the $B=0$ spin-flip cotunneling rate (for $k_\mathrm{B} T>\sqrt{2}t$ and $|\epsilon|<|\Delta|,|e|V_\mathrm{SD}$) is:
\begin{equation}\label{eq:sf-cotunneling}
 W_\mathrm{cot}^0 = \frac{k_\mathrm{B} T}{\pi\hbar} \left[\left(\frac{\hbar\Gamma_\mathrm{S}}{\Delta}\right)^2+\left(\frac{\hbar\Gamma_\mathrm{D}}{\Delta-2 U'-2|e|V_{SD}}\right)^2\right]
\end{equation}
with mutual (inter-dot) charging energy $U'$ and $\Delta = \alpha_1 \delta V_\mathrm{P1}+\alpha_2 \delta V_\mathrm{P2}$ for plunger gate voltages $\delta V_\mathrm{P1,P2}$ measured from the effective $(0,1)-(1,1)-(0,2)$ triple point (lower-left corner of the bias triangle in Fig.~\ref{fig3}(b)). Eq.~\eqref{eq:sf-cotunneling} accounts for virtual transitions between effective $(1,1)$ and $(0,1)$ (first term) as well as effective $(1,1)$ and $(1,2)$ charge states (second term).

In the present case, $\Delta \simeq |e|V_\mathrm{SD}\gg U'$. The higher current level in the upper right corner of Fig.~\ref{fig3}(b) further suggests $\Gamma_{\mathrm{D}}\gg \Gamma_{\mathrm{S}}$, giving (for this particular experiment):
\begin{equation}\label{eq:WcotSimplified}
 W_\mathrm{cot}^0 \simeq \frac{k_\mathrm{B} T}{\pi\hbar} \left(\frac{\hbar\Gamma_\mathrm{D}}{|e|V_\mathrm{SD}}\right)^2.
\end{equation}
Using the above expression for $W_\mathrm{cot}^0$, we then use Eq. \eqref{eq:IvsB} to fit to the $I_\mathrm{SD}(B)$ data in Fig.~\ref{fig5}(b), giving us $\Gamma_{\mathrm{D}}$ = 34 $\mu$eV for the tunneling rate and $T$ = 115 mK for the electron temperature.

The $B=0$ spin-flip cotunneling rate $W_\mathrm{cot}^0$ is energy-independent in the limit $\sqrt{2} t<k_\mathbf{B} T$. However, the leakage current does acquire a dependence on the energy detuning, $\epsilon = \alpha_1 V_\mathrm{P1}-\alpha_2 V_\mathrm{P2}$, when the escape rate from the double-dot due to resonant tunneling is suppressed below the spin-flip cotunneling rate. This leads to a Lorentzian dependence of the current on detuning $\epsilon$ with a $t$-dependent width $\delta\epsilon$:
\begin{eqnarray}\label{eq:Ivsepsilon}
 I_\mathrm{SD}(\epsilon) & = & e\frac{4}{3}\frac{W_\mathrm{cot}^0}{1+\left(\epsilon/\delta\epsilon\right)^2},\quad B=0,\\
 \delta\epsilon &= &\left(\frac{3\Gamma_Dt^2}{W_\mathrm{cot}^0}\right)^{1/2}.
\end{eqnarray}
Eq. (\ref{eq:Ivsepsilon}) is valid in the same limit ($\sqrt{2}t < k_\mathrm{B} T$, $W_\mathrm{cot}^0\ll \Gamma_{\mathrm{S},\mathrm{D}}$) as Eq. (\eqref{eq:IvsB}). In the strong-tunneling limit, $\sqrt{2} t>k_\mathrm{B} T$, the theory predicts that $I(\epsilon)$ should show a strong resonant-tunneling peak of width $\sim t$, followed by a slowly-varying Lorentzian background described by Eq. (\eqref{eq:Ivsepsilon}) at large $\epsilon$. The absence of a strong resonant-tunneling peak in the data of Fig.~\ref{fig5}(c) confirms that the device is operating in the regime $\sqrt{2} t < k_\mathrm{B} T$, justifying our use of Eqs. \eqref{eq:IvsB} and \eqref{eq:Ivsepsilon} to analyse the data.

A nonlinear fit to the $I_\mathrm{SD}(\epsilon)$ data in Figs.~\ref{fig5}(c) using Eq. \eqref{eq:Ivsepsilon} yields $t$ = 0.5 $\mu$eV for the inter-dot tunneling rate, using our previously determined values $\Gamma_{\mathrm{D}}$ = 34 $\mu$eV and $T$ = 115 mK. These parameter values are well within the experimentally expected range. The small value of $t$ indicates weak inter-dot tunnel coupling, consistent with the results shown in Fig.~\ref{fig3}(b). We conclude that the spin-flip cotunneling mechanism provides a fully consistent explanation of the observed leakage current in the spin blockade regime. The mechanism could be applied to reanalyse previous experiments in group IV semiconductors~\cite{Churchill2009} where the nature of the leakage current was not fully understood.

In conclusion, we have presented a lithographically-defined double quantum dot in intrinsic silicon showing excellent charge stability and low disorder. The multi-gate architecture provides independent control of electron number in each dot as well as a tunable tunnel coupling. We observed Pauli spin blockade in an \emph{effective} two-electron system from which we extracted the singlet$-$triplet splitting. The leakage current in the spin blockade regime is well explained by a spin-flip cotunneling mechanism, which could be of widespread importance in group-IV materials with weak hyperfine coupling. The results obtained here provide a pathway towards investigation of spin blockade in silicon double quantum dots with \emph{true} (1,1) and (2,0) electron states. Towards this end, we are planning future experiments incorporating a charge sensor to monitor the last few electrons~\cite{Simmons2007}. We anticipate that such an architecture will provide excellent prospects for realizing singlet$-$triplet qubits in silicon~\cite{Culcer2009}.

\section{Methods}

{\bf Fabrication steps}. The devices investigated in this work were fabricated on a 10~k$\Omega$-cm $n-$type high resistivity $\langle$100$\rangle$ silicon wafer using standard micro-fabrication techniques. The $n^+$ source and drain ohmic contacts regions in Fig.~\ref{fig1}(b) were produced via high concentration phosphorus diffusion at $\sim 1000 \,^{\circ}{\rm C}$, resulting in peak dopant densities of $\sim 10^{20}$~cm$^{-3}$. Next, the high-quality SiO$_2$ of 10 nm thickness was grown via dry thermal oxidation in the central region at 800$\,^{\circ}{\rm C}$ in O$_2$ and dichloroethylene. The barrier gates were first patterned on the thin SiO$_2$ region using electron beam lithography (EBL) followed by thermal evaporation of 40~nm thick aluminium and lift-off process. Before the next EBL step, the barrier gates are exposed to air for 10 mins at 150$\,^{\circ}{\rm C}$ to form $\sim$4~nm of Al$_2$O$_3$ acting as a dielectric layer. This process was repeated for lead gates and plunger gates layers with aluminium thicknesses of 40~nm and 120~nm respectively. A final forming gas anneal (95{\%} N$_2$ and 5{\%} H$_2$) was performed for 15 mins to achieve a low density of Si-SiO$_2$ interface traps, of order 10$^{10}$~cm$^{-2}$ eV$^{-1}$, as measured on a similarly processed chip~\cite{Johnson2010}. The low trap density is clearly reflected in the device stability and the low level of disorder observed in the transport data shown in the results section.

{\bf Experimental setup}. Electrical transport measurements were carried out in a dilution refrigerator with a base temperature $T \sim$ 40 mK. We simultaneously measured both the DC current and the differential conductance $dI$/$dV_{\rm sd}$, the latter using a source-drain AC excitation voltage of 100~$\mu$V at 87 Hz.

\section{Acknowledgements}

The authors thank D. Barber and R. P. Starrett for their technical support and acknowledge the infrastructure support provided by the Australian National Fabrication Facility. This work was funded by the Australian Research Council, the Australian Government, and by the U. S. National Security Agency and U.S. Army Research Office (under Contract No. W911NF-08-1-0527). W.A.C acknowledges the funding from the CIFAR JFA. F.Q. acknowledges funding from NSERC, WIN and QuantumWorks.

\section{Author Contributions}

N.S.L. fabricated the devices. N.S.L., W.H.L and C.H.Y designed and performed the experiments. W.A.C and F.Q. modelled the spin-flip cotunneling rate. N.S.L., F.A.Z., W.A.C, F.Q., A.M. and A.S.D. wrote the manuscript. A.S.D planned the project. All authors discussed the results and commented on the manuscript at all stages.

\end{document}